# 80-Gbit/s 100-m Free-Space Optical Data Transmission Link via a Flying UAV Using Multiplexing of Orbital-Angular-Momentum Beams

Long Li[1]*†, Runzhou Zhang[1]†, Zhe Zhao[1], Guodong Xie[1], Peicheng Liao[1], Kai Pang[1], Haoqian Song[1], Cong Liu[1], Yongxiong Ren[1], Guillaume Labroille[2], Pu Jian[2], Dmitry Starodubov[1], Robert Bock[3], Moshe Tur[4], Alan E. Willner[1]*

**Affiliations:**

[1]Department of Electrical Engineering, University of Southern California, Los Angeles, California 90089, USA.

[2]CAILabs, Rennes 35200, France.

[3]R-DEX Systems, Marietta, Georgia 30068, USA.

[4]School of Electrical Engineering, Tel Aviv University, Ramat Aviv 69978, Israel.

*Correspondence to: longl@usc.edu, willner@usc.edu.

†These authors contributed equally to this work.

**Abstract**: We explore the use of orbital-angular-momentum (OAM)-multiplexing to increase the capacity of free-space data transmission to moving platforms, with an added potential benefit of decreasing the probability of data intercept. Specifically, we experimentally demonstrate and characterize the performance of an OAM-multiplexed, free-space optical (FSO) communications link between a ground station and a moving unmanned-aerial-vehicle (UAV). We achieve a total capacity of 80 Gbit/s up to 100-m-roundtrip link by multiplexing 2 OAM beams, each carrying a 40-Gbit/s quadrature-phase-shift-keying (QPSK) signal. Moreover, we investigate for static, hovering, and moving conditions the effects of channel impairments, including: tracking errors, propeller-induced airflows, power loss, intermodal crosstalk, and system bit error rate (BER). We find the following: (a) when the UAV hovers in the air, the power on the desired mode fluctuates by 2.1 dB, while the crosstalk to the other mode is -19 dB below the power on the desired mode; and (b) when the UAV moves in the air, the power fluctuation on the desired mode increases to 4.3 dB and the crosstalk to the other mode increases to -10 dB. Furthermore, the channel crosstalk decreases with an increase in OAM mode spacing.

**One Sentence Summary:** We experimentally demonstrate an OAM-multiplexed 80-Gbit/s free-space optical communication 100-m-roundtrip link between a flying UAV and ground station.

The data communications capacity needs of manned and unmanned aerial platforms have been increasing dramatically over the past several years, thereby driving the need for higher-capacity links between these platforms and their ground stations (*1-4*). One example of an aerial platform is an unmanned-aerial-vehicle (UAV), such as flying drones that are proliferating for numerous applications (*5-8*). In addition to the need for high-speed communications, there is also the desire to minimize the probability of possible interception of the data exchange in order to achieve enhanced privacy and security (*9-11*).

Due to the higher carrier frequency of the lightwave, FSO communications generally holds the promise of having both higher capacity and lower probability of intercept (LPI) than radio-frequency (RF) and millimeter-wave techniques (*12-14*). Specifically, there have been several reports of FSO communication links with moving aerial platforms (*15-17*).

Importantly, an approach for significantly increasing capacity for fixed ground-based FSO links has gained interest over the past few years. This technique, known as space-division-multiplexing (SDM), is based on the simultaneous transmission of multiple independent data-carrying beams (*18*). Mode-division-multiplexing (MDM) is a subset of SDM, where each of the multiple beams is a unique mode from an orthogonal modal basis set (*19-21*). Orthogonality minimizes crosstalk among the modes and enables efficient multiplexing at the transmitter, co-propagation of overlapping beams, and low-crosstalk demultiplexing at the receiver (*22*). Although different orthogonal modal basis sets could be used, one possibility is to use OAM modes that are conveniently circularly symmetric (*23, 24*).

OAM modes are characterized by a phase front having an angular dependence of the form $\exp(il\varphi)$, where $\varphi$ is the azimuthal angle and $\ell$ is the OAM order and counts the number of $2\pi$ phase shifts in the azimuthal direction. $\ell$ is an integer which can assume a positive, negative, or zero value corresponding to a clockwise phase helicity, counter-clockwise phase helicity, or no helicity (i.e., a conventional Gaussian beam), respectively. These helical beams "twist" as they propagate (*23*). The intensity of an OAM beam with non-zero order (e.g., Laguerre-Gaussian beam with non-zero azimuthal order and zero radial order) is circularly symmetric and has a ring shape with little power in the center (*23, 24*).

Prior art in FSO communications with multiple OAM beams between two fixed ground stations includes: (a) 2.5 Tbit/s over ~1 m using 32 OAM modes (*25*); (b) 100 Tbit/s over ~1 m using 1008 OAM modes comprised of 12 OAM modes on each of 2 polarizations and 42 wavelengths (*26*); (c) 80 Gbit/s over 260 m using 2 OAM modes and 400 Gbit/s over 120 m using 4 modes (*27,28*); (d) <1-kbit/s single-beam transmission of OAM superpositions between two 143-km-spaced islands (*29*).

To date, there has been little reported on the use of OAM multiplexing in FSO communications between a ground station and a moving platform. Importantly, such a scenario is likely to face a few major challenges arising from the special structured nature of the OAM beams themselves. Challenge include the following: (a) ***Alignment***: Low inherent crosstalk and power-coupling loss generally relies on accurate on-axis detection of the multiple OAM beams (*30*), thereby necessitating more tracking sophistication for an OAM-multiplexed link over a single conventional Gaussian-based link; (b) ***Turbulence***: Turbulence resulting from the atmosphere or from a UAV's propellers could significantly distort the OAM beam's phase front, thus resulting in increased received power fluctuations and channel crosstalk, as compared to recovering a single conventional Gaussian beam (*31,32*).

In this paper, we explore the use of OAM multiplexing to increase the capacity and decrease the probability of intercept of data transmission to moving platforms. We experimentally investigate the performance of an FSO communication link between a ground station and a flying UAV transmitting two multiplexed OAM beams up to ~100 m roundtrip distance. For ease of demonstration, our UAV carries a retro-reflector but does not carry a transmitter or receiver. Instead, the ground station transmits the beams to the UAV, whereupon the beams are reflected back to a receiver on the ground station that is co-located with the transmitter (see experimental setup). Each OAM beam carries a 40-Gbit/s QPSK signal, thereby a total capacity of 80 Gbit/s at a single carrier wavelength of 1550 nm is achieved. We measure the impact of channel impairments, including tracking errors and propeller-induced airflow on beam quality and system performance, in terms of received signal power, intermodal crosstalk among channels, and BERs. We find that: (a) when the UAV hovers in the air, the power on the desired mode fluctuates by 2.1 dB over a 60-second period, whereas the crosstalk to the other mode is -19 dB below the power on the desired mode, and the crosstalk fluctuates within a 7.8-dB range; and (b) when the UAV moves in the air at a speed of ~0.1 m/s, the power fluctuation on the desired mode increases to 4.3 dB and the crosstalk to the other mode increases to -10 dB below the power on the desired mode. Furthermore, the channel crosstalk decreases with an increase in OAM mode spacing, and the number of error-free transmitted data frames increases when channel OAM mode spacing increases from 2 to 3.

**OAM-multiplexing in ground-to-UAV FSO communications**

Figure 1 illustrates our prospective application for using OAM multiplexing in high-capacity FSO communications between a UAV and a ground station. An experimental schematic of our specific link is shown in Fig. 2A. The ground station contains an OAM transmitter (Tx), an OAM receiver (Rx), and a beam tracking system. A retro-reflector carried by the UAV is flown up to ~50 m away (i.e., ~100 m round trip) from the ground station to efficiently reflect the OAM beams coming from the transmitter back to the receiver with little distortion. During the experiment, the octocopter UAV moves and hovers at different locations up to ~20 m above the ground and up to ~50 m away (see Supplementary Section 1 for the experimental environment).

The OAM transmitter optics are shown in Fig. 2B. We use a custom-designed OAM (de)multiplexer pair based on multi-plane mode conversion for: (a) generating and multiplexing multiple OAM beams at the Tx, and (b) demultiplexing and receiving them at the Rx (*33*). The OAM multiplexer has seven single-mode-fiber (SMF) pig-tailed input ports, in which different Gaussian beams from different inputs are converted to co-axially propagating OAM beams; the seven input ports correspond to OAM $\ell$ = -3, -2, -1, 0, +1, +2, +3. In a back-to-back measurement of an OAM (de)multiplexer pair, the highest power-coupling loss for a desired mode is ~11.8 dB, and the highest crosstalk is ~-19.1 dB from a mode to its nearest mode (i.e., mode spacing of 1), ~-23.9 dB from a mode to its second nearest mode (i.e., mode spacing of 2), and ~-26.1 dB from a mode to its third nearest mode (i.e., mode spacing of 3) (see Supplementary Section 2 for measurements of the OAM (de)multiplexer). Such crosstalk level could be sufficient to enable error-less high speed communications (*34*).

A 20-Gbaud (40-Gbit/s) QPSK data signal at 1550 nm is amplified by an Erbium-doped fiber amplifier (EDFA) and split into two beams, one of which is delayed using a ~10-m SMF to decorrelate the data sequence (*35*). These two beams are sent to two of the seven input ports of the OAM multiplexer, generating two multiplexed OAM beams. Simultaneously, a 1530-nm probe beam used for tracking is sent to the $\ell$ = 0 input of the OAM multiplexer. These co-axially propagating beams are transmitted through a 1:10 beam expander to enlarge the beam sizes, after which they propagate in free-space to the gimbal-mounted retro-reflector on the UAV. The diameter of the transmitted beams are ~3 cm for Gaussian, ~4.2 cm for OAM ±1, ~5.2 cm for OAM ±2, and ~6 cm for OAM ±3 beams. The retro-reflector reverses the order of an OAM beam between +$\ell$ and -$\ell$, and the relative purity of the OAM beam itself is not significantly altered (*36*, see Supplementary Section 3 for measurements of OAM beam quality reflected from the retro-reflector).

The tracking system points the expanded beams to the retro-reflector (see Supplementary Section 4 for details of the beam tracking system). The gimbal ensures that the retro-reflector faces the ground station, such that the reflected beams can be received (see Supplementary Section 5 for specifications of the UAV and the gimbal). The diameter of the received beams are ~3.1 cm for Gaussian, ~4.3 cm for OAM ±1, ~5.3 cm for OAM ±2, and ~6.2 cm for OAM ±3 beams. Since the Tx and Rx share the same aperture which has a diameter of ~7.1 cm, the forward path and the return path coincide with each other and are separated using a 4-inch 50:50 beamsplitter. After beam reduction by an 8:1 beam reducer, the beams are demultiplexed for coherent optical detection; coherent detection uses a local oscillator to track the phase of the received beam, thus recovering both the phase and intensity information carried by the received beam (*37*, see Supplementary Section 6 for details of the optical signal generation and coherent detection).

**Potential challenges for system performance**

In general, beam tracking is considered important for single-beam non-OAM FSO links due to the relatively small beam diameters (*1-8*, *15-17*). This issue is even more pronounced for OAM-multiplexed systems, since any deviation from coaxial detection of the uniquely-structured beams can produce both additional power-coupling loss of the desired mode and crosstalk from one mode to other unwanted modes, as shown in Figs. 3E and 3F (*30*). In ideal cases, the Tx and Rx are perfectly aligned, as shown in Fig. 3A. However, in a dynamic ground-to-UAV FSO link, the Tx and Rx may have residual tracking accuracy limitations. Such limitations that may result in residual tracking errors could lead to various misalignment problems, including lateral displacement and tip/tilt error at the Tx and Rx, as shown in Figs. 3B-3D (*30*). As only one example in our experiment, the retro-reflector itself can produce two misalignment issues: (a) It reflects the beams back in their original direction within an error of <1 arcsecond due to fabrication imperfections, which produces a ~0.2 mm offset and is a relatively minor issue at a ~100-m-roundtrip distance; and (b) Unless the beams are exactly in the center of the retro-reflector (which was not perfectly the case in our system), there is a natural lateral displacement such that the Rx center does not completely overlap with the reflected beams' center.

In order to determine the sensitivity of our system to misalignment, we transmit one mode and measure the received power on different OAM modes for various displacements between the Rx center and the axis of the received beam. For these initial results: (a) the UAV and the tracking system are turned off, (b) the retro-reflector on the UAV is placed on the ground ~50 m away from the transceiver, and (c) the lateral deviation of the beam away from the center of the retro-reflector that produces a lateral displacement between the center of the receiver and the axis of the received beam is varied by manually changing the beam location. Figures 3G and 3H show the measured received power on different OAM modes under various horizontal and vertical displacements. Furthermore, simulation results for our system are shown in Fig. 3I (*30*), which show similar trends of the dependence of the received power on displacement. However, the magnitude of the crosstalk to the wrong mode is experimentally higher partially due to the non-ideal performance of the OAM (de)multiplexer (see Supplementary Section 2 for measurements of the OAM (de)multiplexers). Moreover, our system is more tolerant to vertical as compared to horizontal displacement, which is likely due to the fact that the OAM demultiplexer has an OAM-dependent transfer function along the horizontal direction (*33*). Our measurement indicates that a horizontal displacement of >3 mm would lead to a power-coupling loss of >1.5 dB for a desired mode and a crosstalk of >-13 dB between two OAM modes with a mode spacing of 4; this crosstalk might be considered high for >40-Gbit/s QPSK signal transmissions (*34*); we note that such a displacement (i.e., 30-μrad tracking error at 100 m) is readily mitigated by tracking systems (*1-8*, *15-17*). Finally, Figs. 3G-3I also show that smaller OAM spacing would lead to higher crosstalk under similar displacement conditions. Misalignment arising from mechanical UAV vibrations, should also be mitigated by the tracking system.

Importantly, the tight alignment tolerance for low crosstalk and low power loss in an OAM-multiplexed link can be viewed as a potential benefit of increasing the difficulty of eavesdropping (i.e., LPI) by any off-axis receiver (*30*). As shown in Fig. 3H when comparing a single Gaussian beam to 2 multiplexed OAM beams, a displacement of 4 mm of an off-axis eavesdropper when transmitting OAM -1 and -3 simultaneously would result in: (a) an increased power loss of ~2.5 dB, and (b) a crosstalk level between -1 and -3 modes is ~ -4.2 dB, whereas crosstalk is not an issue for eavesdropping a Gaussian beam link.

Another technical challenge for even single Gaussian beam systems is atmospheric turbulence. For an OAM-multiplexed link, such turbulence would dynamically distort the structured phase front of the beams. Since the power coupling is mode specific and the orthogonality of multiple co-propagating OAM beams depends on their unique helical phase-fronts, this distortion would cause received power fluctuations of the desired mode and increased channel crosstalk from the unwanted modes (*31*, *32*). Additionally, the airflow induced by the UAV propellers may distort the OAM beam's phase front and result in increased received power fluctuation and channel crosstalk. To estimate this effect, we measure the Rytov variance ($\sigma^2$) using a transmitted 1550-nm Gaussian probe beam and a ~1-mm diameter point detector as outlined in Ref. (*38*) over a 10-minute period. In order to separate mechanical vibrations from airflow turbulence, the retro-reflector is detached from and placed underneath the UAV. The retro-reflector is fixed ~50 m away from the transceiver. The ~5-cm diameter probe beam propagates from the transmitter to the retro-reflector and back to the receiver. We compare the measured power distributions when the UAV's propellers are off or on as shown in Figs. 4A and 4B, respectively. By fitting the power distribution into a lognormal distribution (*39, 40*), we find that the $\sigma^2$ value is ~0.0028 and the refractive-index structure parameter ($C_n^2$) is ~$9.4 \times 10^{-15}$ m$^{-2/3}$ when the propeller is turned off. Due to the increased airflow when the UAV propellers are turned on, $\sigma^2$ increases to ~0.0080.

In our communication system, the aperture is large enough to capture the entire beam and the received power may benefit from the aperture averaging effect (*38*). To verify this, we measure the received OAM spectrum when OAM +1 beam is transmitted for the scenarios of the propellers off and on, as shown in Figs. 4C and 4D, respectively. We observe that the overall impact of the propellers on the received OAM spectrum does not appear to be significant in our measurement.

**System performance measurements under flying conditions**

The system measurements for the OAM-multiplexed FSO link between the flying (i.e., hovering or moving) UAV and the ground transceiver are performed under clear weather conditions in the daytime. The wind varies but is typically ~8 km/h from west to east, and the UAV is located ~50 m northwest of the ground transceiver.

In order to evaluate the effects of beam jitter caused by various issues (including those outlined in the previous section), we measure the statistics of the received beam centroid before beam reduction. Figures 5A-5C show the distributions of relative position of beam centroid when the UAV is static on the ground with tracking system off, hovers in the air ~10 m above ground with tracking system on, moves horizontally in the air at a speed of ~0.1 m/s with tracking system on, respectively. OAM +3 beam is transmitted during the measurements. The statistics of each scenario is obtained by continuously capture 1000 intensity profiles of the beam over a 120-second period using an infrared camera. The beam jitter variance is ~0.0218 mm$^2$ when the UAV is grounded, and it increases to ~0.0877 mm$^2$ and ~0.4604 mm$^2$ when the UAV is hovering and moving, respectively. We believe that these increases are caused by the accuracy and speed limitations of our self-made tracking system, which can be significantly improved by advanced tracking systems (*15-17, 38*).

Figure 5D shows different UAV positions, and Figs. 5E-5I show the coupled signal power and modal crosstalk under hovering and moving conditions. Figure 5E shows the received power on different modes during a continuous 60-second period when OAM +1 beam is transmitted and the UAV is hovering ~10 m above ground and ~50 m away from the ground station (location 1 in Fig. 5D). The power on the desired mode (i.e., OAM +1) fluctuates within a 2.1-dB range. The crosstalk on the OAM +3 mode is <-19 dB below the power coupled into the desired mode, and the crosstalk fluctuates within a 7.8-dB range. Figures 5F-5I show the measured OAM spectrum under different flight conditions when OAM +1 is transmitted, and the shaded portion of each bar shows the power fluctuation range. We observe that both the power on the desired mode and the crosstalk into other modes experience more fluctuations when the UAV is moving than when it is hovering, which agrees with our beam jitter measurements of Figs. 5A-5C.

To verify link performance under flying conditions, Fig. 6 shows measurements of the BER for each 4096-symbol data frame. OAM +3 and OAM -1 beams are multiplexed and transmitted simultaneously, each carrying a 40-Gbit/s QPSK signal. Figure 6A shows a 60-second time sequence of BER measurements of the OAM +3 channel when the UAV is hovering in location 1 of Fig. 5D. The transmitted power of each channel is fixed at 10 dBm. We observe that the BERs fluctuate, mostly being below (or even error free) but sometimes above the 7% forward-error correction (FEC) limit (*41*), we again note that a better tracking system would significantly improve link performance.

Figure 6B shows BER measurements as a function of transmitted channel power, such that there is a range of measurement values at each power level due to the non-ideal beam tracking. For each transmitted channel power level, 10 BER measurements (i.e., 10 data frames) are taking within a ~30-second measurement time period. The value at the bottom of the figure represents the number of error-free frames out of the 10 measured frames for each power level.

Since crosstalk tends to be higher for neighboring modes than for those far away (see Figs. 3G-3I) (*30, 31*), larger channel mode spacing could produce lower crosstalk and potentially better link performance. Figure 6C shows the BER measurements for different transmitted channel mode spacing when the UAV is hovering at location 1 of Fig. 5D. The results indicate that the performance is better for mode spacing of 3 than for 2. As an example, the number of error-free frames increases from 4 to 7 when mode spacing increases from 2 to 3 when transmitted power is 7 dBm. Figure 6D shows the BER for OAM +3 when OAM +3 and -1 are transmitted for different link distances and when the UAV is hovering. A slight decrease on the number of error-free frames is observed when the transmission distance increases from 40 m to 100 m, which may indicate a minor performance degradation.

**Discussion**

This paper describes our demonstration of an 80-Gbit/s FSO communication link multiplexing 2 OAM modes between a UAV and a ground station up to a 100 m roundtrip distance. Although our UAV only carries a reflector and acts as a simple relay for the ground transceiver, we believe that our results likely validate that a communication link with a UAV that carries transmitter/receiver equipment can be realized. Moreover, with the addressed challenges properly met, we believe that future high-capacity communication links to moving platforms enabled by OAM multiplexing has the potential to achieve multi-Tbit/s capacities over multi-km distances.

To extend the transmission distance to multi-km, the following issues should be considered:

(a) ***Divergence***: OAM beams have a vortex intensity profile and a beam divergence that grows with $\ell$. To capture sufficient signal power at the receiver or limit the amount of beam divergence, larger receiver aperture sizes or larger transmitted beam sizes than in our experiment would be required at longer distances, respectively. For example, a transmitted OAM +3 beam with a diameter of 20 cm would be ~20 cm at 1 km, ~75 cm at 10 km, and ~7.1 m at 100 km (*30*). With a 20-cm-diameter receiver aperture, the link loss due to beam divergence would be 1.7 dB at 1 km, 27 dB at 10 km, 120 dB at 100 km. In order to achieve a <20-dB power loss due solely to divergence, the receiver aperture diameter should be >23 cm at 10 km, and >2.2 m at 100 km.

(b) ***Turbulence***: The effect of atmospheric turbulence would be more significant as transmission distances increase. For a similar atmospheric condition as in our demonstration (i.e., a $C_n^2$ of ~$9.4 \times 10^{-15}$ m$^{-2/3}$), the $\sigma^2$ would increase to ~0.19 at 1 km and even higher at 10 km (*37*). Although in our experiment we did not use any compensation techniques, adaptive optics (AO), multiple-input-multiple-output (MIMO)-based channel equalization, and other digital-signal-processing (DSP) techniques have been demonstrated to help mitigate turbulence effects in stationary OAM-multiplexed FSO links (*42-47*). As an example, promising turbulence mitigation techniques in OAM links have been experimentally demonstrated for relatively weak turbulence (*42-44*) and theoretically analyzed for moderate-to-strong turbulence (*45-47*). Moreover, much of the turbulence compensation need not necessarily be performed on the UAV but can be performed instead on the ground station as pre- and/or post-mitigation, as was shown using AO for a bi-directional link in Ref. (*42*).

(c) ***Tracking***: The unique structure of OAM beams places a premium on accurate tracking. In general, there have been advances both commercially and experimentally that has the potential to meet the system requirements (*1-8, 15-17*). Furthermore, the vortex amplitude profile has sharp gradients that may actually help the tracking system performance under certain conditions (*48*).

To increase the link capacity to beyond Tbit/s, there exist a number of potential approaches, including the following:

(a) ***Mode Spacing***: More OAM modes can be accommodated in a given link by reducing the mode spacing and ensuring low inter-modal crosstalk, and we believe that a mode spacing of 2 for a UAV link is achievable. Of course, optical components (e.g., (de)multiplexer) with still higher performance would be helpful. Furthermore, MIMO-based channel equalization techniques can be used to mitigate crosstalk; however, the total number of modes may be limited due to the increased signal processing complexity.

(b) ***Mode Order***: More modes that are located at higher orders can be achieved by utilizing larger optical elements since beam size and beam divergence increase with larger OAM value. For example, with the same transmitted beam size of 20 cm and a distance of 1 km, an OAM beam of +3 and +20 has a diameter at the receiver of ~20 cm and ~47 cm, respectively.

Beyond using only OAM multiplexing, there might well be key advantages to dramatically increasing capacity by employing channel multiplexing in multiple domains, such as wavelength- and polarization-division-multiplexing (WDM and PDM) (*25,26*). Indeed, fixed FSO links have used OAM+WDM+PDM in the lab to achieve 100 Tbit/s (*26*). However, the various optical components should maintain their high performance across the link's wavelength spectrum, and the integrity of the polarization states should be preserved.

Finally, to further enhance the privacy/security offered by the use of OAM beams, mode hopping a data channel's location among different modes can be employed, in analogy with frequency hopping in radio-based communication links (*49, 50*).

**Acknowledgments:**

We thank Dr. Michael Kendra from the Air Force Office of Scientific Research (AFOSR), and Dr. David H. Hughes and Dr. Saba Mudaliar from the Air Force Research Laboratory (AFRL) for their valuable insights and support. We acknowledge the generous support of AFOSR FA9550-16-C-0008; National Science Foundation (NSF) ECCS-1509965 and IIP-1622777; Vannevar Bush Faculty Fellowship program sponsored by the Basic Research Office of the Assistant Secretary of Defense for Research and Engineering and funded by the Office of Naval Research (ONR) through grant N00014-16-1-2813.

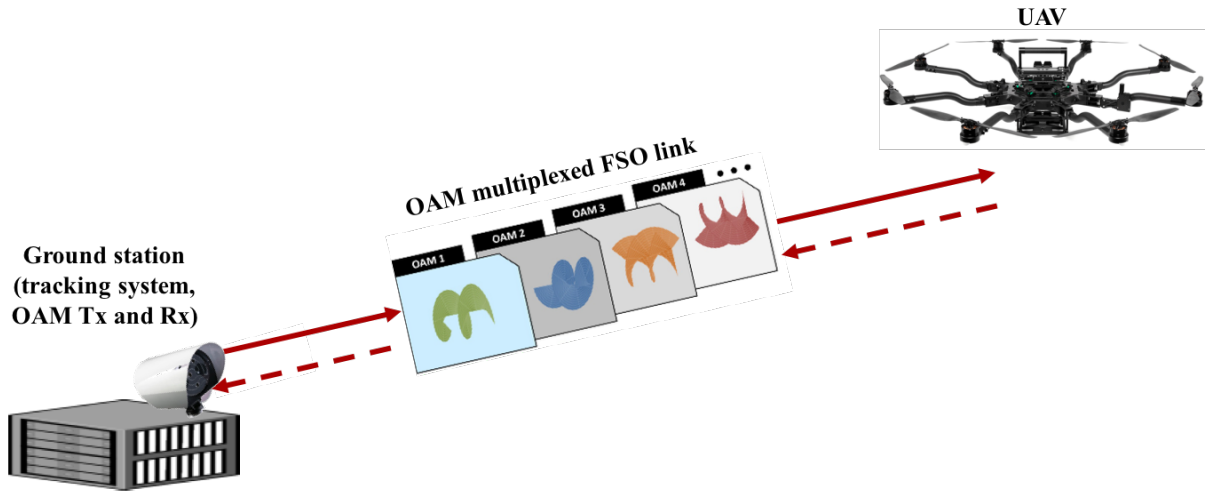

**Fig. 1**. **Concept of a free-space optical (FSO) communication link between an unmanned-aerial-vehicle (UAV) and a ground station using OAM multiplexing.** The ground station includes the tracking system as well as the orbital-angular-momentum (OAM) transmitter and receiver, and the UAV hovers in the air.

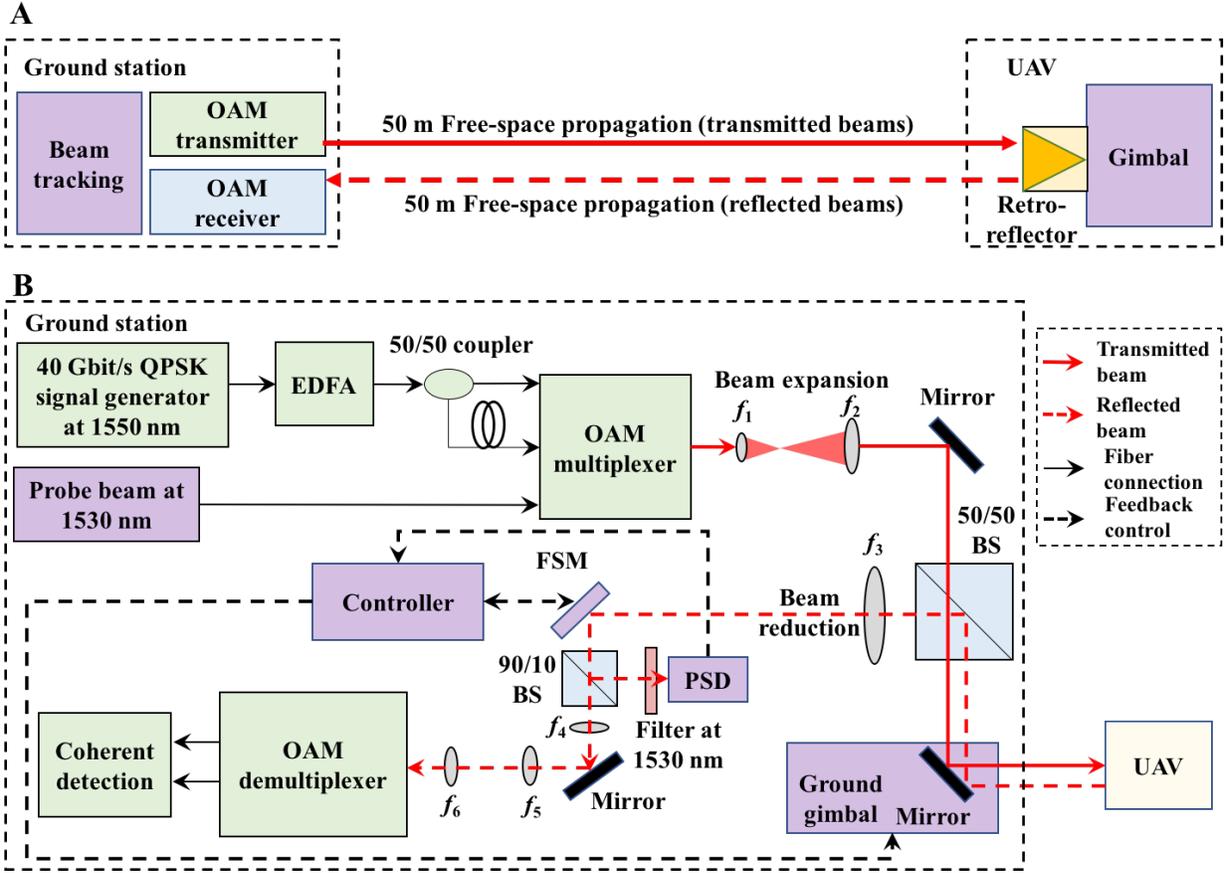

**Fig. 2. Experimental setup of an FSO communication link between a UAV and a ground station using OAM multiplexing.** (**A**) Schematic diagram of the ground station and the UAV. (**B**) Ground station setup, including the OAM transmitter, the OAM receiver and the beam tracking system. BS: beamsplitter; EDFA: Erbium-doped-fiber-amplifier; FSM: fast-steering mirror; FSO: free-space optical; OAM: orbital-angular-momentum; PSD: position sensitive detector; QPSK: quadrature-phase-shift-keying; Rx: receiver; Tx: transmitter; UAV: unmanned-aerial-vehicle. Focal lengths of lenses: $f_1$: 50 mm; $f_2$: 500 mm; $f_3$: 600 mm; $f_4$: 75 mm; $f_5$: 100 mm; $f_6$: 100 mm.

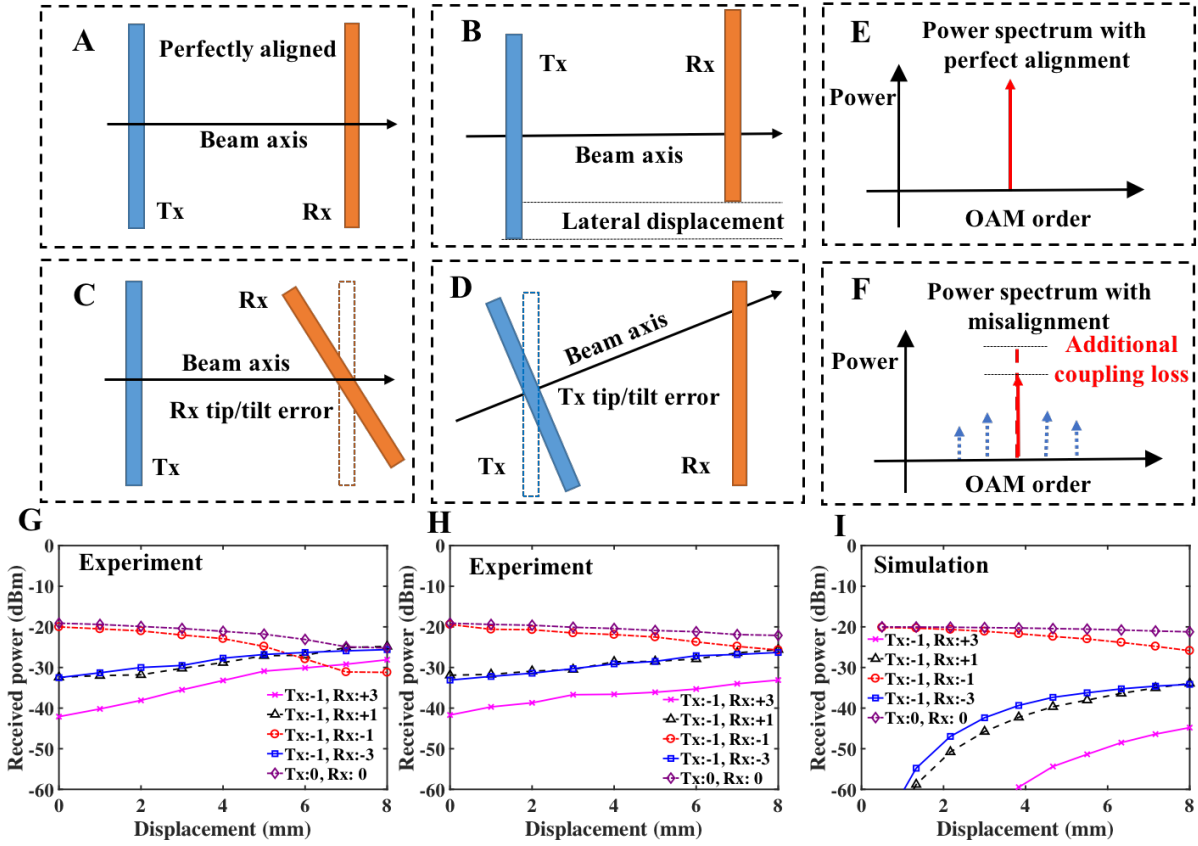

**Fig. 3. Effects of misalignment on system performance.** (**A**)-(**D**) Illustration of OAM-based communications link with perfect alignment, displacement, tip/tilt error at Rx, and tip/tilt error at Tx, respectively. (**E**) and (**F**) Received OAM spectrum with perfect alignment and misalignment between the transmitter and the receiver, respectively. (**G**)-(**I**) Received power on different OAM modes under horizontal, vertical, and simulated displacement when OAM $\ell$=-1 is transmitted. The roundtrip transmission distance is ~100 m and the tracking system is off. The displacement refers to the distance between the OAM beam center and the center of the receiver. Tx:-1, Rx:+3: received power on OAM +3 mode when OAM -1 beam is transmitted.

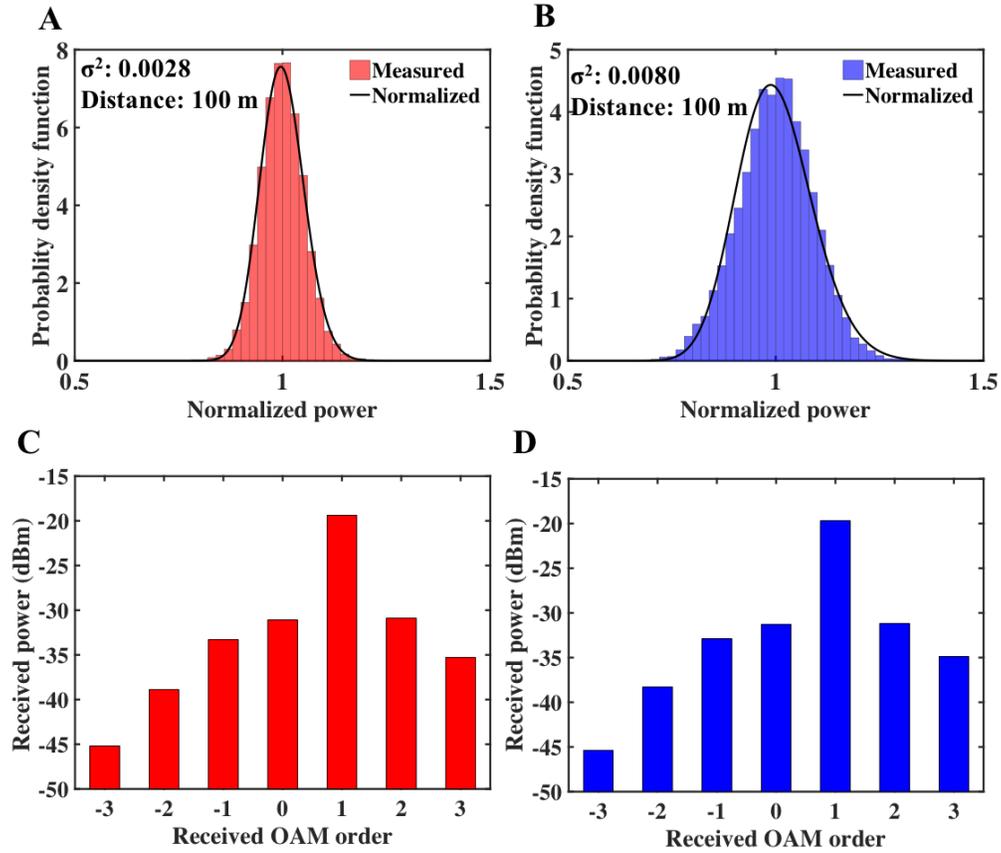

**Fig. 4**. **Effects of propeller-induced airflow on system performance.** (**A**) and (**B**) Measured power distribution when the UAV's propellers are turned off or on, respectively. The power is normalized to its mean during the measurement. (**C**) and (**D**) Measured OAM spectrum when the UAV's propellers are turned off and on, respectively, when OAM $\ell=+1$ is transmitted over ~100 m roundtrip.

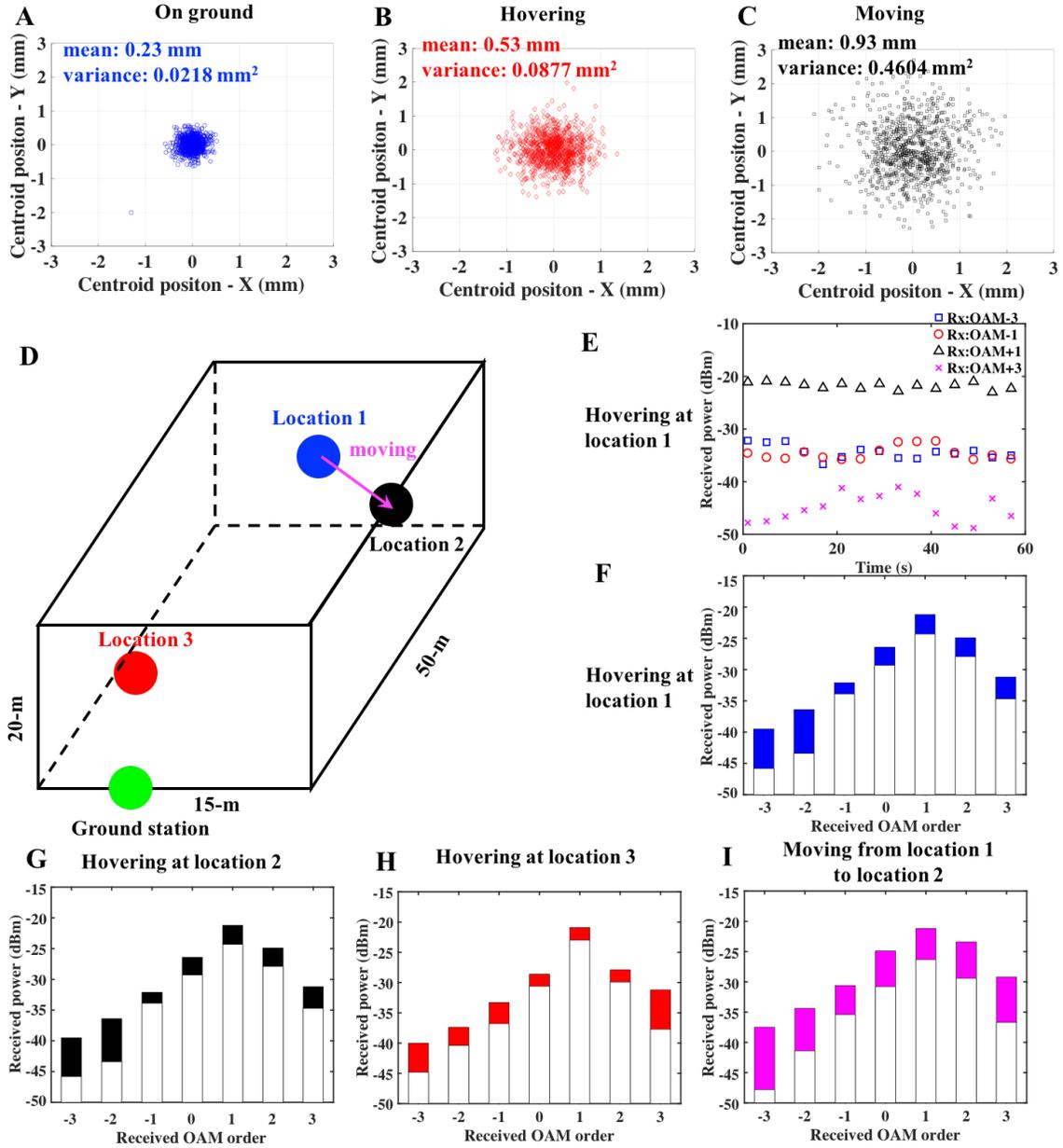

**Fig. 5. Beam jitter, power, and crosstalk measurements in flight environment.** (**A**)-(**C**) Measured statistics of beam displacement with respect to the receiver center when the UAV is static on the ground, hovering at location 1, moving from location 1 to location 2 at a speed of ~0.1 m/s, respectively. (**D**) Schematic diagram of different locations where the UAV hovers. Location 1 is ~10 m above ground, ~50 m away from the transceiver with an angular position of 0°; location 2 is ~20 m above ground, ~40 m away from the transceiver with an angular position of 15°; location 3 is ~5 m above ground, ~10 m away from the transceiver with an angular position of -5°. (**E**) The received power on different modes in a 60-second period, when OAM -1 is transmitted and the UAV is hovering at location 1. (**F**)-(**I**) OAM spectrum in a 60-second period when OAM +1 is transmitted and the UAV is hovering at location 1, location 2, and location 3, and moving from location 1 to location 2 at a speed of ~0.1 m/s, respectively.

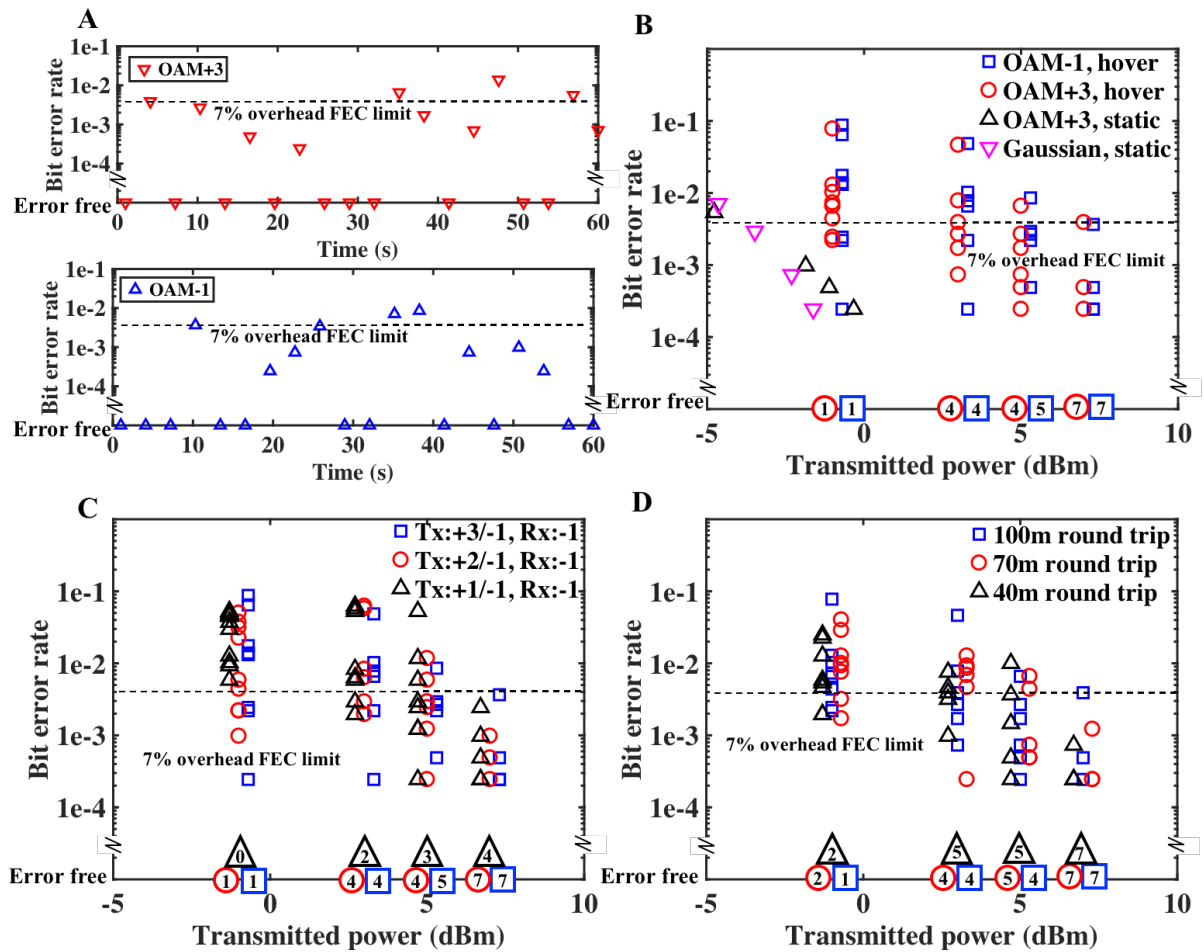

**Fig. 6**. **Bit-error-rate (BER) measurements in flight environment.** Each channel transmits a 40-Gbit/s quadrature phase shift keying (QPSK) signal. (**A**) BERs in a 60-second period and (**B**) BERs measurements for OAM $\ell=+3$ and $\ell=-1$, when OAM $\ell=+3$ and $\ell=-1$ beams are transmitted. At each transmitted power level, 10 data frames (each has 4096 bits) over ~30-second period are measured. The number at the bottom represents the number of error-free frames during the measurement period at a certain transmitted power level. (**C**) BER measurements when two OAM beams with different mode spacing are transmitted. The UAV is hovering ~10 m above the ground and ~50 m away from the ground station. (**D**) BER measurements for OAM +3 when OAM +3 and -1 are transmitted. The UAV is hovering at different distances away from the ground station. FEC: forward-error-correction.

**Supplementary Materials:**

Materials and Methods

Figures S1-S5

**Supplementary Materials:**

**Materials and Methods:**

**1. Experimental environment:**

Our experiment is carried out on a lawn, as shown in Fig. S1A. A ground station, where the OAM transmitter, OAM receiver, and beam tracking system, integrated on an optical table, is placed at one side of the lawn, as shown in Fig. S1B. A retro-reflector carried by a UAV is placed up on the other side of the lawn, as Fig. S1C shows, up to ~50 m away from the ground station to reflect the OAM beams coming from the transmitter back to the receiver. The UAV, which carries the gimbal-mounted retro-reflector. During the test it starts from the ground, climbs up, moves to certain locations, hovers at certain locations, and lands.

**2. Measurements of the OAM (de)multiplexers:**

The structure and concept of the 7-mode OAM (de)multiplexer is illustrated in Fig. S2A. The successive phase profiles for multi-plane mode conversions are all printed on a single reflective phase plate, with each phase profile being located on a different phase spot of the phase plate. After successively projecting multiple Gaussian (or OAM) beams to those phase spots, multiple OAM channels could be (de)multiplexed.

The intensity profiles of the generated Gaussian beam and OAM beams are shown in Fig. S2B. The diameter of the transmitted beams are ~3 mm for Gaussian, ~4.2 mm for OAM ±1, ~5.2 mm for OAM ±2, and ~6 mm for OAM ±3 beams. We measure the the power coupling loss and crosstalk for all modes, when a fixed (de)multiplexer pair is well aligned and separated by ~1 m. As shown in Fig. S2C, the highest power-coupling loss for a desired mode is ~11.8 dB, and the highest crosstalk is ~-19.1 dB from a mode to its nearest mode (i.e., mode spacing of 1), ~-23.9 dB from a mode to its second nearest mode (i.e., mode spacing of 2), and ~-26.1 dB from a mode to its third nearest mode (i.e., mode spacing of 3).

**3. Measurements of OAM beam quality reflected from the retro-reflector:**

We measure the effects of retro-reflector on the OAM beams quality. The experimental setup is shown in Fig. S3A. An OAM beam at 1550 nm is generated by launching a collimated Gaussian beam with a beam waist of 2.2 mm onto a spatial light modulator (SLM) loaded with a specific pattern. The generated OAM beam is then reflected by a retro-reflector with a diameter of 2 inches. We measure the OAM power spectrum of the reflected OAM beam (shown in Figs. S2B-S2D) and the transmitted OAM beam (shown in Fig. S3E), and we observe that the OAM order of the reflected beam is the negative of that of the incoming beam, but its OAM power spectrum shows similar power coupled into the desired mode as well as crosstalk to the neighboring modes.

**4. Details of the beam tracking system:**

The beam tracking system consists of a fine tracking subsystem and a coarse tracking subsystem, as shown in Fig. S4A and S4B. The fine tracking system consists of a fast-steering mirror (FSM), a position sensitive detector (PSD) and a feedback controller. The PSD detects the relative position of the reflected Gaussian beam relative to the its center, and sends this position information to the controller so that the FSM could be adjusted accordingly to keep the reflected OAM beams propagating to the center of the PSD, thus achieving accurate beam tracking. The FSM has a 2-inch diameter, a <2-µrad angular resolution, a 750-Hz tuning rate, and a ±1.5° tuning range. The control algorithm of the feedback controller is integrated onto a circuit board, and analog signals are used for FSM tuning.

The coarse tracking system consists of a rotation stage, a goniometer, an angular position sensor and a feedback controller. The angular position sensor detects the angular position of the FSM of the fine tracking system, and sends it to the controller. The rotation stage and the goniometer of the coarse tracking system are adjusted accordingly when the FSM of the fine tracking system is out of its operation range. The goniometer is mounted on top of the rotation stage, which could tune the beams vertically and horizontally, respectively. The rotation stage and the goniometer have a $>\pm 45°$ tuning range in both directions, with a ~20-μrad angular resolution, and a 15°/sec maximum tuning speed. The mirror mounted on the rotation stage and goniometer is 10 cm in diameter. The control algorithm uses digital signals to tune the rotation stage and the goniometer, and the time consumption in one feedback loop is about 33 ms.

### 5. Specifications of the UAV and the gimbal:

The retro-reflector is mounted on a gimbal and carried by an octocopter UAV. The gimbal targets the retro-reflector to the ground platform using GPS information with an accuracy of ~2-m, i.e., ~12° at a distance of 10 m, and 1.2° at a distance of 100 m. The angular range of the incoming beam that our retro-reflector could reflect is between ~$\pm 15°$ referring to its axis. Therefore this targeting accuracy would satisfy the requirements of our tracking system. The gross weight of the retro-reflector is 4 kg, which is within the maximum payload of the gimbal (5.5 kg). The gross weight of the gimbal is 1.5 kg, thus a total weight of 5.5 kg is carried by the UAV, which is also within its maximum payload of 9 kg.

### 6. Details of the optical signal generation and coherent detection:

The schematic diagram of the indoor optical modules is shown in Fig. S5. A 1550-nm light is generated by a laser source and modulated by a Lithium Niobate (LiNbO$_3$) In-phase/Quadrature (I/Q) modulator. A polarization controller (PC) is used to control the polarization of the input light to match with that of the I/Q modulator. A 20-GHz clock source is used to generate the 20-Gbaud (i.e., 40-Gbit/s) QPSK signal with a pseudorandom binary sequence (PRBS) of $2^{31}-1$. After being amplified by the Erbium-doped fiber amplifier (EDFA), the modulated light is sent to the on field OAM-multiplexed FSO communications link.

For coherent optical detection, the received signal is amplified by an EDFA, and the optical signal detector (homodyne receiver) is used to detect the received signal, and calculate its bit-error-rate (BER).

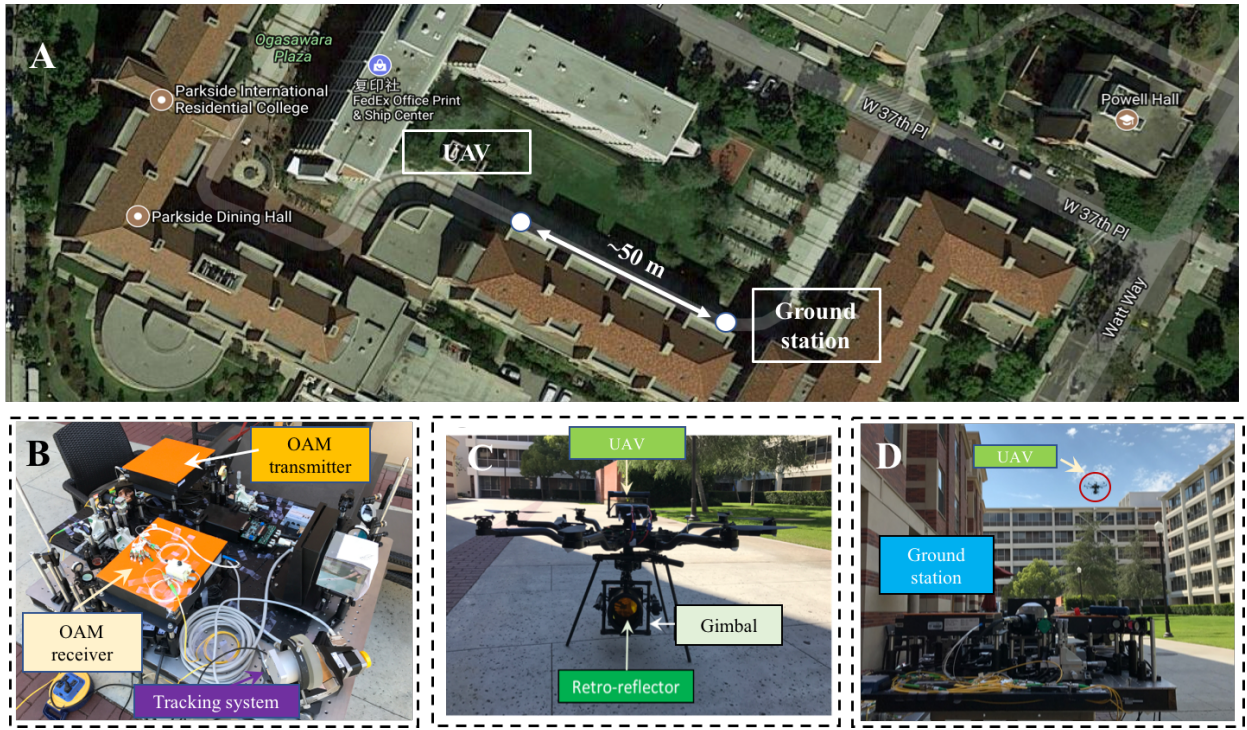

**Fig. S1**. **Photos of experimental setup.** (**A**) Location of the test field. (**B**) Ground station on the test field. (**C**) UAV on the test field; (**D**) Communication between the ground station and the hovering UAV;

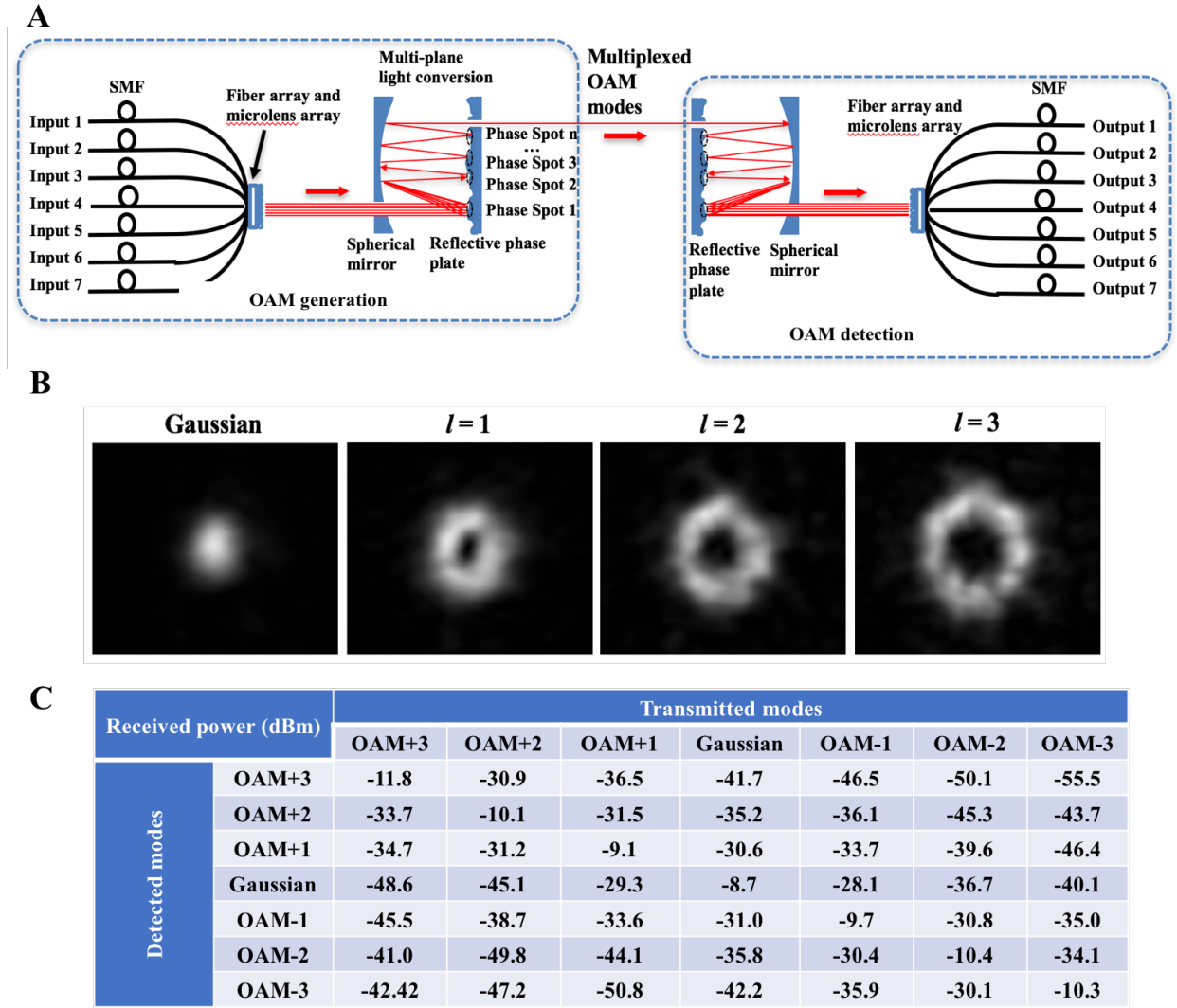

**Fig. S2**. **Concept and Measurements of the OAM (de)multiplexers.** (**A**) Structure and concept of the 7-mode OAM (de)multiplexer. (**B**) Intensity profiles of the generated Gaussian beam (i.e., $\ell=0$) and OAM beams (i.e., $\ell=+1,+2,+3$). (C) Measured power coupling loss and crosstalk for all modes in a fixed ~1-m back-to-back link between an OAM (de)multiplexer pair with good alignment. The input power to each input port of the OAM multiplexer is 0 dBm.
| Received power (dBm) | | Transmitted modes | | | | | | |
|---|---|---|---|---|---|---|---|---|
| | | OAM+3 | OAM+2 | OAM+1 | Gaussian | OAM-1 | OAM-2 | OAM-3 |
| Detected modes | OAM+3 | -11.8 | -30.9 | -36.5 | -41.7 | -46.5 | -50.1 | -55.5 |
| | OAM+2 | -33.7 | -10.1 | -31.5 | -35.2 | -36.1 | -45.3 | -43.7 |
| | OAM+1 | -34.7 | -31.2 | -9.1 | -30.6 | -33.7 | -39.6 | -46.4 |
| | Gaussian | -48.6 | -45.1 | -29.3 | -8.7 | -28.1 | -36.7 | -40.1 |
| | OAM-1 | -45.5 | -38.7 | -33.6 | -31.0 | -9.7 | -30.8 | -35.0 |
| | OAM-2 | -41.0 | -49.8 | -44.1 | -35.8 | -30.4 | -10.4 | -34.1 |
| | OAM-3 | -42.42 | -47.2 | -50.8 | -42.2 | -35.9 | -30.1 | -10.3 |
21

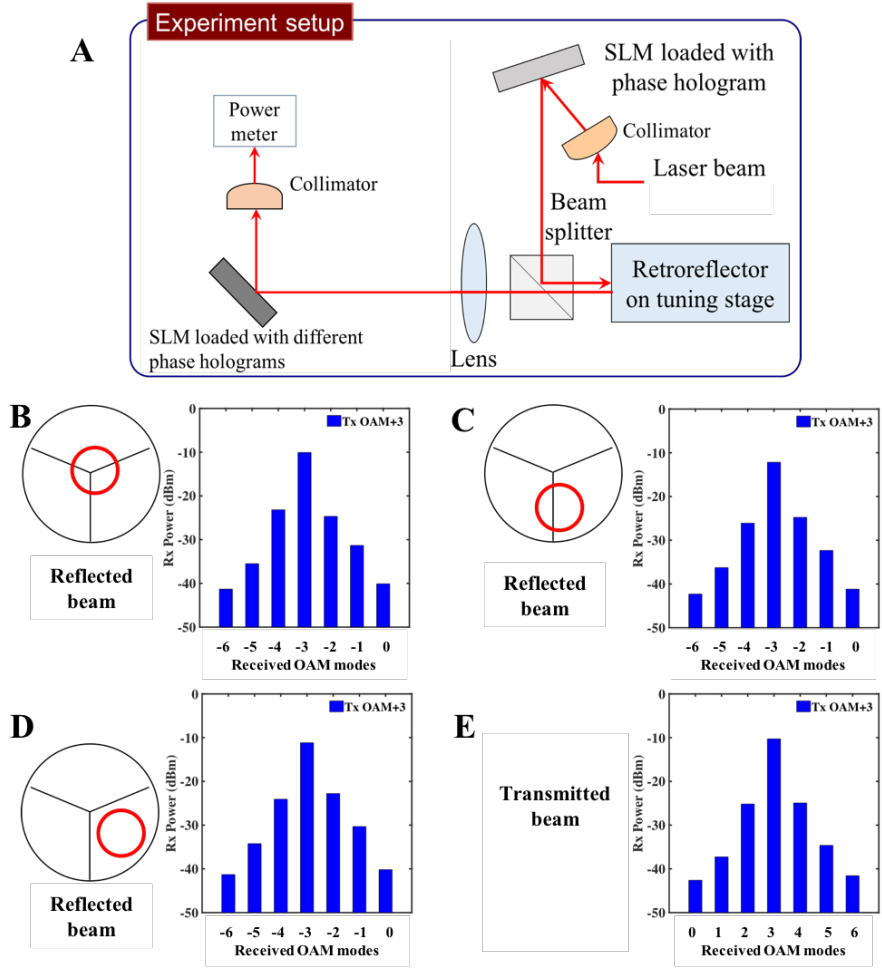

**Fig. S3**. **Measurements on the effect of retro-reflector on the OAM beams.** (**A**) Experimental setup. (**B**)-(**E**) Measured OAM power spectrum of reflected OAM beam and the transmitted OAM beam. OAM +3 beam is transmitted launched on different part of the retro-reflector. The retro-reflector has a diameter of 2 inches. After reflection, a beam's OAM order changes between $+\ell$ and $-\ell$.

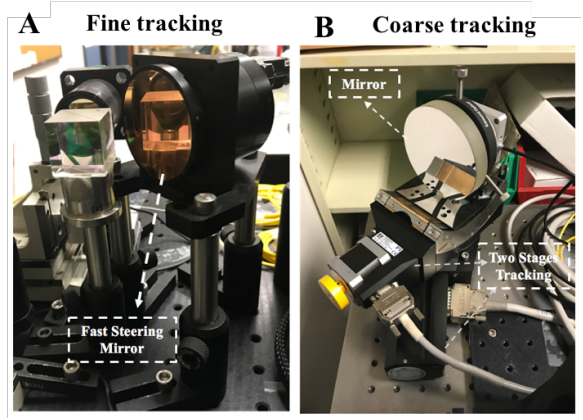

**Fig. S4**. **Photos of the beam tracking system.** (**A**) The fast steering mirror (FSM) and the position detector (PSD) of the fine tracking system. (**B**) The rotation stage, the goniometer and the mirror of the coarse tracking system.

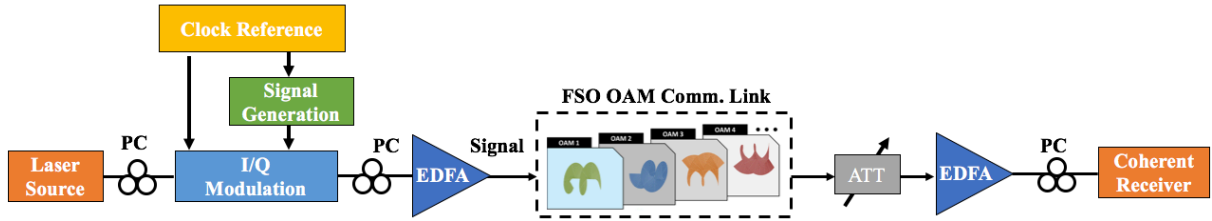

**Fig. S5**. **Schematic diagram of signal generation and coherent optical detection.** ATT: optical attenuator; Comm.: communications; EDFA: Erbium-doped fiber amplifier; FSO: free-space optical; I/Q: in-phase and quadrature; OAM: orbital angular momentum; PC: polarization controller.